\documentclass[11pt]{article}
\usepackage{graphicx}
\usepackage{epsfig}

\newcommand{\BABARPubYear}    {03}

\newcommand{\BABARConfNumber} {023}
\newcommand{\SLACPubNumber} {10093}

\long\def\inst#1{\par\nobreak\kern 4pt\nobreak
    {\it #1}\par\vskip 10pt plus 3pt minus 3pt}

\RequirePackage{xspace}
\usepackage{relsize}
\def\qqbar {\ensuremath{q\overline q}\xspace}
\def\babar{\mbox{\slshape B\kern-0.1em{\smaller A}\kern-0.1em
    B\kern-0.1em{\smaller A\kern-0.2em R}}}
\def\Bbar    {\kern 0.18em\overline{\kern -0.18em B}{}\xspace}

\def\BB      {\ensuremath{B\Bbar}\xspace} 
\def\Bz      {\ensuremath{B^0}\xspace}
\def\Bzb     {\ensuremath{\Bbar^0}\xspace}
\def\BzBzb   {\ensuremath{\Bz {\kern -0.16em \Bzb}}\xspace}
\def\Bu      {\ensuremath{B^+}\xspace}
\def\Bub     {\ensuremath{B^-}\xspace}

\def\BpBm    {\ensuremath{\Bu {\kern -0.16em \Bub}}\xspace}

\newcommand{\optbar}[1]{\shortstack{{\tiny (\rule[.4ex]{1em}{.1mm})}
  \\ [-.7ex] $#1$}}
\def\BorBbar    {\kern 0.18em\optbar{\kern -0.18em B}{}\xspace}
\def\DorDbar    {\kern 0.18em\optbar{\kern -0.18em D}{}\xspace}
\def\KorKbar    {\kern 0.18em\optbar{\kern -0.18em K}{}\xspace}
\def\CP                {\ensuremath{C\!P}\xspace}
\def\pep2{PEP-II}
\mathchardef\Upsilon="7107
\def\Y#1S{\ensuremath{\Upsilon{(#1S)}}\xspace}% no space before {...}!

\def\FourS {\Y4S}

\begin{document}
{\pagestyle{empty}

\begin{flushright}
\babar-CONF-\BABARPubYear/\BABARConfNumber \\
SLAC-PUB-\SLACPubNumber \\
%hep-ex/\LANLNumber \\
August 2003 \\
\end{flushright}

%\par\vskip 5cm
\par\vskip 3cm

% Title of the paper
\begin{center}
\Large \bf\boldmath Observation of the Decay $B^0\to\rho^+\rho^-$ and\\ 
Measurement of the Branching Fraction and Polarization  
\end{center}
\bigskip

\begin{center}
\large The \babar\ Collaboration\\
\mbox{ }\\
August 9, 2003
%\today
\end{center}
\bigskip \bigskip

% Abstract
\begin{center}
\large \bf Abstract
\end{center}
We have observed the rare decay $B^0\to\rho^+\rho^-$
in a sample of 89 million $\BB$ pairs recorded 
with the $\babar$ detector.
We measure the branching fraction
${\cal B}(B^0\to\rho^+\rho^-)=(27^{+7+5}_{-6-7})\times 10^{-6}$
and determine the longitudinal polarization fraction
${\Gamma_L}/\Gamma=0.99^{+0.01}_{-0.07}\pm 0.03$. 
Our results are preliminary.

\vfill
\begin{center}
Contributed to the  
XXI$^{\rm st}$ International Symposium on Lepton and Photon Interactions at High~Energies, 8/11 --- 8/16/2003, Fermilab, Illinois USA
\end{center}

\vspace{1.0cm}
\begin{center}
{\em Stanford Linear Accelerator Center, Stanford University, 
Stanford, CA 94309} \\ \vspace{0.1cm}\hrule\vspace{0.1cm}
Work supported in part by Department of Energy contract DE-AC03-76SF00515.
\end{center}

\newpage
} % end of pagestyle{empty}

\begin{center}
\small

The \babar\ Collaboration,
\bigskip

%% author list as of 02-Jun-2003 (595 authors)
%
B.~Aubert,
R.~Barate,
D.~Boutigny,
J.-M.~Gaillard,
A.~Hicheur,
Y.~Karyotakis,
J.~P.~Lees,
P.~Robbe,
V.~Tisserand,
A.~Zghiche
\inst{Laboratoire de Physique des Particules, F-74941 Annecy-le-Vieux, France }
% ...

\end{center}\newpage

%%%%%%%%%%%%%%%%%%%%%%%%%%%%%%%%%%%%%%%%%%%%%%%%%%%%%%%%%%%%%%%%%%%%%%%%%
% INTRODUCTION
%%%%%%%%%%%%%%%%%%%%%%%%%%%%%%%%%%%%%%%%%%%%%%%%%%%%%%%%%%%%%%%%%%%%%%%%%

\section{INTRODUCTION}
\label{sec:Introduction}

Charmless $B$ meson decays provide an opportunity to measure 
the weak-interaction phases arising from the elements of the 
Cabibbo-Kobayashi-Maskawa (CKM) quark-mixing matrix~\cite{Kobayashi}.
There has been increasing interest in the study
of $B\to\pi\pi$ and $\rho\pi$ decays where the time-dependent
$\CP$-violating asymmetries are related to CKM angle
$\alpha\equiv {\rm arg}\,[\, - V^{ }_{td}V^*_{tb}\,/
\,V^{}_{ud}V^*_{ub}\,]$
and interference between the tree and
penguin amplitudes could give rise to direct-$\CP$ violation.
The decay\footnote{Charge conjugation 
is implied here and throughout this paper unless explicitly stated.}
$B^0\to\rho^+\rho^-$ is another promising
mode for $\CP$ violation studies and constraints on
the angle $\alpha$.
The angular correlations in this decay to two vector particles
introduce additional complications in the analysis, but the
measurement of the magnitudes or phases of the helicity amplitudes
will provide better understanding of the decay 
models~\cite{bvv1, bvv2, bvv3}.

\begin{figure}[htbp]
\begin{center}
\centerline{
\epsfig{figure=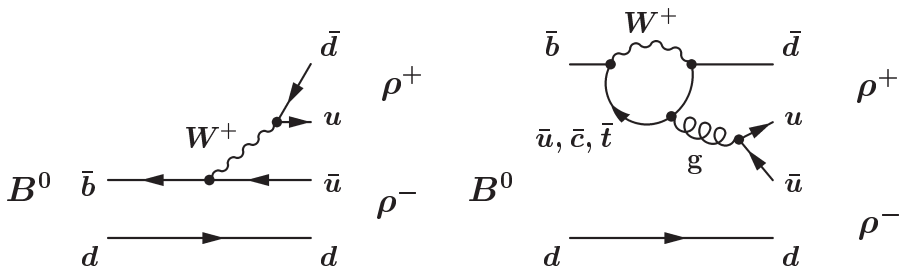,height=1.5in}
}
\caption{\sl 
Two of the diagrams describing the decays 
$B^0\to\rho^+\rho^-$.
}
\label{fig:Diagram}
\end{center}
\end{figure}

The decay $B^0\to\rho^+\rho^-$ 
is expected to proceed through tree-level $b\to u$ 
transitions and CKM-suppressed $b\to d$ penguins
as illustrated in Fig.~\ref{fig:Diagram}. 
%%%%%%%%%%%%%%%%%%%%%%%%%%%%%%%%%%%%%%
%%%%%%%%%%%%%%%%%%%%%%%%%%%%%%%%%%%%%%
The presence of penguins and both 
$\CP$-even (S- and D-wave) and $\CP$-odd (P-wave)
components in the decay amplitude complicate 
the measurement of $\alpha$.
Isospin relations among the three
$B\to\rho\rho$ modes may reduce the uncertainties 
in the measurement of $\alpha$ due to penguin 
contributions (penguin pollution), analogous
to the methods proposed for $B\to\pi\pi$~\cite{grossmanquinn}.
The recent limit on the $B^0\to\rho^0\rho^0$ 
decay rate~\cite{babarVV} and the measurements of the
$B^+\to\rho^+\rho^0$ branching fraction~\cite{babarVV, belle} 
result in experimental limits on the amount of penguin pollution, 
while $B^+\to\rho^+\rho^0$ polarization measurements 
provide evidence that the $\CP$-even longitudinal component 
dominates in the $B\to\rho\rho$ decay amplitudes.
In this paper we report on the observation of the 
$B^0\to\rho^+\rho^-$ decay mode and the measurement
of its branching fraction and the longitudinal polarization
fraction in the decay.

%%%%%%%%%%%%%%%%%%%%%%%%%%%%%%%%%%%%%%%%%%%%%%%%%%%%%%%%%%%%%%%%%%%%%%%%%
% THE BABAR DETECTOR AND DATASET
%%%%%%%%%%%%%%%%%%%%%%%%%%%%%%%%%%%%%%%%%%%%%%%%%%%%%%%%%%%%%%%%%%%%%%%%%

\section{THE \babar\ DETECTOR AND DATASET}
\label{sec:babar}

In this analysis, we use the data collected with the 
\babar\ detector~\cite{babar} at the \pep2 asymmetric-energy 
$e^+e^-$ collider~\cite{pep} operated at the 
center-of-mass (CM) energy of the $\FourS$ resonance
($\sqrt{s}=10.58$~GeV).
These data represent an integrated luminosity 
of 81.9~fb$^{-1}$, corresponding to 88.9 million 
$\BB$ pairs, at the $\FourS$ energy 
(on-resonance) and 9.6~fb$^{-1}$ approximately 
40~MeV below this energy (off-resonance). 

Charged-particle momenta are measured in a tracking system 
consisting a five-layer double-sided silicon vertex tracker (SVT) 
and a 40-layer central drift chamber (DCH), 
both immersed in a 1.5 T axial magnetic field. 
\babar\ achieves an impact parameter resolution
of about 40~$\mu$m for the high momentum charged particles
in the $B$ decay final states, allowing the precise determination 
of decay vertices.
The tracking system covers 92\% of the solid angle in the CM frame.

Charged-particle identification is provided by 
measurements of the energy loss 
(${\rm d}E/{\rm d}x$) in the tracking devices (SVT and DCH) and
by an internally reflecting ring-imaging Cherenkov detector 
(DIRC) covering the central region. 
A $K$-$\pi$ separation of 
better than four standard deviations ($\sigma$) is achieved for 
momenta below 3~GeV/$c$, decreasing to 2.5~$\sigma$ at the highest 
momenta in the $B$ decay final states.
Photons are detected by a CsI(Tl) electromagnetic calorimeter
(EMC). The EMC provides good
energy and angular resolution for detection of photons 
in the range from 20~MeV to 4~GeV.
The energy and angular resolutions are $3\%$ and 4 mrad, 
respectively, for a 1~GeV photon.

%%%%%%%%%%%%%%%%%%%%%%%%%%%%%%%%%%%%%%%%%%%%%%%%%%%%%%%%%%%%%%%%%%%%%%
% EVENT SELECTION
%%%%%%%%%%%%%%%%%%%%%%%%%%%%%%%%%%%%%%%%%%%%%%%%%%%%%%%%%%%%%%%%%%%%%%

\section{EVENT SELECTION}
\label{sec:Selection}

Hadronic events are selected based on track multiplicity and 
event topology. We fully reconstruct $B^0\to\rho^+\rho^-$ 
candidates from their decay products 
$\rho^\pm\to\pi^\pm\pi^0$ and $\pi^0\rightarrow \gamma\gamma$.
Charged track candidates are required to originate 
from the interaction point, and to have at least 12 DCH hits 
and a minimum transverse momentum of 0.1~GeV/$c$. 
Charged pion tracks are distinguished from kaon and proton tracks 
via a likelihood ratio that includes 
${\rm d}E/{\rm d}x$ information from the SVT and DCH
for momenta below 0.7~GeV/$c$ and
the DIRC Cherenkov angle and number of photons
for higher momenta.
They are also distinguished from electrons primarily on 
the basis of the EMC shower energy and lateral shower moments.

\begin{figure}[htbp]
\begin{center}
\centerline{
\epsfig{figure=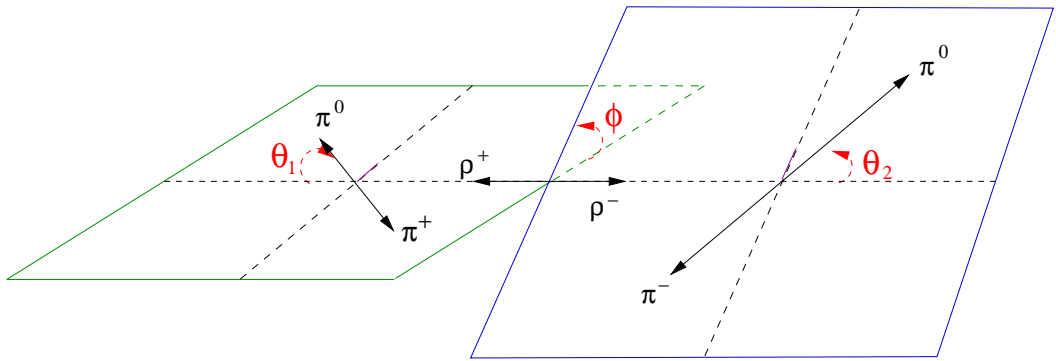,width=5.5in}
}
\caption{\sl 
Definition of helicity angles ($\theta_1$ and $\theta_2$)
in ${B^0\to\rho^+\rho^-}$ decays.
}
\label{fig:helangles}
\end{center}
\end{figure}

We reconstruct $\pi^0$ mesons from pairs of photons, each with 
a minimum energy of 30 MeV, with the 
shower shape consistent with the photon hypothesis,
and not matched to a track.
The typical width of the reconstructed $\pi^0$ mass is 7~MeV. 
We accept $\pi^0$ candidates in the invariant mass
interval $\pm$15~MeV from the nominal mass.
We select $\rho$ candidates to
satisfy $0.52~{\rm GeV} < m_{\pi\pi} <1.02$~GeV.
The helicity angles $\theta_{1}$ and $\theta_{2}$
of $\rho^+$ and $\rho^-$
are defined as the angles between the $\pi^0$ direction
in the $\rho$ rest frame and the direction
of the $\rho$ boost with respect to the $B$
as shown in Fig.~\ref{fig:helangles}.
We restrict the helicity angles 
to the region $-0.75\le\cos\theta_{1,2}\le0.95$
to suppress combinatorial background and 
reduce acceptance uncertainties due to 
low-momentum pion reconstruction.

%%%%%%%%%%%%%%%%%%%%%%%%%%%%%%%%%%%%%%
%%%%%%%%%%%%%%%%%%%%%%%%%%%%%%%%%%%%%%

To reject the dominant continuum background 
(from $e^+e^-\to\qqbar$ events, $q = u, d, s, c$),
we require $|\cos\theta_T| < 0.8$, where $\theta_T$ 
is the angle between the thrust axis of the $B$ candidate 
and the thrust axis
of the rest of the tracks and neutral clusters in
the event, calculated in the CM frame. 
The distribution of $|\cos{\theta_T}|$ is 
sharply peaked near $1$ for jet-like events originating 
from $\qqbar$ pairs and nearly uniform for the 
isotropic decays of the $B$ meson.
We also construct a Fisher discriminant (${\cal F}$)
that combines 11 variables: 
the polar angle of the $B$ momentum vector, 
the polar angle of the $B$-candidate thrust axis,
both calculated with respect to the beam axis in the CM frame,
and the scalar sum of the CM momenta of charged particles
and photons (excluding particles from the $B$ candidate)
entering nine coaxial angular intervals of 10$^\circ$
around the $B$-candidate thrust axis~\cite{CLEO-fisher}.

We identify $B$ meson candidates using two nearly 
independent kinematic variables \cite{babar},
the beam energy-substituted mass $m_{\rm{ES}} =$ 
$[{ (s/2 + \mathbf{p}_i \cdot \mathbf{p}_B)^2 / E_i^2 - 
\mathbf{p}_B^{\,2} }]^{1/2}$ and the energy difference
$\Delta E = (E_i E_B - \mathbf{p}_i 
\cdot \mathbf{p}_B - s/2)/\sqrt{s}\,$,
where $(E_i,\mathbf{p}_i)$ is the $e^+e^-$ initial state 
four-momentum, and $(E_B,\mathbf{p}_B)$
is the four-momentum of the reconstructed $B$ candidate,
all defined in the laboratory frame.
For signal events the $m_{\rm{ES}}$ distribution 
peaks at the $B$ mass and the $\Delta E$ distribution 
peaks near zero.
Our initial selection requires $m_{\rm{ES}}>5.2$ GeV
and $|\Delta E|<0.2$~GeV, while the signal resolution
is roughly 3~MeV and 50~MeV, respectively. 
The selected sample contains 54042 events most of which 
are retained in sidebands of the variables for later fitting.

%%%%%%%%%%%%%%%%%%%%%%%%%%%%%%%%%%%%%%%%%%%%%%%%%%%%%%%%%%%%%%%%%%%%%%%%%
% ANALYSIS METHOD
%%%%%%%%%%%%%%%%%%%%%%%%%%%%%%%%%%%%%%%%%%%%%%%%%%%%%%%%%%%%%%%%%%%%%%%%%

\section{ANALYSIS METHOD}
\label{sec:Analysis}

We use an unbinned, extended maximum-likelihood (ML) fit to extract
simultaneously the signal yield and angular polarization. 
There are three event yield ($n_{j}$) categories $j$: signal, 
continuum~$\qqbar$, and $\BB$ combinatorial background.
We define the likelihood for each $B^0\to\rho^+\rho^-$
event candidate~$i$:
%%%%%%%%%%%%%%%%%%%%%%%
\begin{equation}
{\cal L}_i = \sum_{j=1}^{3} n_{j}\, 
{\cal P}_{j}(\vec{x}_{i};\vec{\alpha}) ,
\label{eq:likelev}
\end{equation}
%%%%%%%%%%%%%%%%%%%%%%%
where each of the ${\cal P}_{j}(\vec{x}_{i};\vec{\alpha})$ is 
the probability density function (PDF) for measured variables 
$\vec{x}_{i}$.
The numbers $\vec{\alpha}$ parameterize the expected 
PDFs of measured variables in each category.
We allow for multiple candidates in a given event by assigning 
to each selected candidate a weight of $1/N_i\,$, 
where $N_i$ is the number of candidates in that same event.
The average number of candidates per event is 1.27.
The extended likelihood for a sample 
of $N_{\rm cand}$ candidates is 
%%%%%%%%%%%%%%%%%%%%%%%
\begin{equation}
{\cal L} = \exp\left(-\sum_{j=1}^{3} n_{j}\right)\, 
\prod_{i=1}^{N_{\rm cand}} 
\exp\left(\frac{\ln{\cal L}_i}{N_i}\right) .
\label{eq:likel}
\end{equation}
%%%%%%%%%%%%%%%%%%%%%%%

The seven fit input variables $\vec{x}_{i}$ are $m_{\rm{ES}}$,
$\Delta E$, ${\cal F}$, invariant masses 
of the $\rho^+$ and $\rho^-$ candidates, and the
corresponding helicity angles $\theta_{\rm 1}$ and 
$\theta_{\rm 2}$.
The correlations among the fit input variables 
for the data and signal Monte Carlo (MC)~\cite{geant} 
are found to be small (typically less than $5\%$), 
except for angular 
correlations in the signal as discussed below.
The PDF, ${\cal P}_{j}(\vec{x}_{i};\vec{\alpha})$,
for a given candidate $i$ is the product of those
for each of the variables and a joint PDF for the  
helicity angles, which accounts for the angular correlations 
in the signal and for detector acceptance effects.
We integrate over the angle $\phi$ between the two 
decay planes shown in~Fig.~\ref{fig:helangles}, 
leaving a PDF that depends 
only on the two helicity angles and the unknown longitudinal 
polarization fraction $f_L\equiv{\Gamma_L}/{\Gamma}$,
where $\Gamma_L$ and ${\Gamma}$ are the 
longitudinal and total decay widths.
The differential decay width~\cite{bvv1} is defined as:
%%%%%%%%%%%%%%%%%%%%%%%
\begin{eqnarray}
{1 \over \Gamma} \  {{\rm d}^2\Gamma \over 
{\rm d}\cos \theta_1 \, {\rm d}\cos \theta_2} = 
{9 \over 4} \left \{ {1 \over 4} (1 - f_L)
\sin^2 \theta_1 \sin^2 \theta_2 + f_L \cos^2 \theta_1 \cos^2 \theta_2 \right\} \ .
\label{eq:helicityintegr}
\end{eqnarray}
%%%%%%%%%%%%%%%%%%%%%%%

The PDF parameters $\vec{\alpha}$, except for $f_L$,
are extracted from MC simulation and on-resonance 
$m_{\rm{ES}}$ and $\Delta E$ sidebands.
The MC resolutions are adjusted by comparisons of data and simulation 
in abundant calibration channels with similar kinematics and topology,
such as $B\rightarrow D\pi, D\rho$ with $D\rightarrow K\pi\pi, K\pi$.
To describe the signal distributions, we use Gaussian functions 
for the parameterization of the PDFs for $m_{\rm{ES}}$ and
$\Delta E$, and a relativistic P-wave Breit-Wigner distribution
for the $\rho$ resonance masses. The angular acceptance effects
are parameterized with empirical polynomial functions for each
helicity angle and are included in the joint helicity-angle PDF as 
a product with ideal distribution in Eq.~(\ref{eq:helicityintegr}).
For the background PDFs, we use polynomials or, in the 
case of $m_{\rm{ES}}$, an empirical phase-space function~\cite{argus}.
In the background PDF we incorporate a small linear 
correlation between the curvature of the phase-space 
function and the event shape variable ${\cal F}$.
The background parameterizations for the $\rho$ candidate masses 
also include a resonant component to account for $\rho$ 
production in the continuum. 
The background helicity-angle distribution is 
also separated into contributions from combinatorial background
and from real $\rho$ mesons, both described by polynomials.
The PDF for ${\cal F}$ 
is represented by a Gaussian distribution 
with different widths above and below the mean
for both signal and background.

There is a fraction of incorrectly reconstructed (fake)
$B^0\to\rho^+\rho^-$ events expected in the selected sample.
This happens when at least one candidate photon in a $\pi^0$ 
candidate or one charged track in a $\rho$ candidate 
belongs to the decay tree of the other $B$. MC simulation 
shows that about 30\% of selected $B^0\to\rho^+\rho^-$ 
events with longitudinal polarization do not have the 
correct decay tree reconstructed, while about 20\% of the 
events have both correctly and incorrectly reconstructed 
decay candidates.
We do not account explicitly for the fake events in the signal 
PDF parameterization since they introduce substantial tails 
in the distributions; these tails have weak discrimination power 
from background and we cannot fully rely on the MC simulation
of the fake component distributions. We account for these effects 
in the signal efficiency evaluation.

MC simulation indicates that about $5\%$ of the events 
in the final sample are from other $B$ decays.  
This background, arising mainly from $b\to c$ transitions, 
is explicitly accounted for in the fit.  
PDFs for this background are taken from MC including a contribution 
from charmless decays such as $B\to\rho\pi$, $\rho^0\rho^+$, 
$\rho K^*$, $a_1\pi$, and $a_1\rho$. The branching fractions 
for these and many other exclusive modes were taken from 
the most recent experimental measurements~\cite{pdg} or
extrapolated from other results with flavor symmetry
approximation. Their contribution was shown 
to be well accounted for by a single $B$-background 
fit component. 
In this analysis we do not explicitly include 
a fit component for other partial waves, including 
non-resonant decays, with the same final-state particles 
selected within the $\rho$ resonance mass window.
These types of decays are assumed to be negligible, 
they are significantly suppressed by the mass
and helicity-angle information in the fit, and they are
examined in the mass and helicity-angle distributions as
discussed below. 

The event yields $n_j$ and polarization $f_L$ are obtained by 
minimizing the quantity $\chi^2\equiv -2\ln{\cal L}$. 
The dependence of $\chi^2$ on a fit parameter
$n_j$ or $f_L$ is obtained with the other
fit parameters floating.
We quote statistical errors corresponding to a unit
increase in $\chi^2$.
The statistical significance of the signal is defined as the 
square root of the change in $\chi^2$ when constraining 
the number of signal events to zero in the likelihood fit.

%%%%%%%%%%%%%%%%%%%%%%%%%%%%%%%%%%%%%%%%%%%%%%%%%%%%%%%%%%%%%%%%%%%%%%%%%
% PHYSICS RESULTS
%%%%%%%%%%%%%%%%%%%%%%%%%%%%%%%%%%%%%%%%%%%%%%%%%%%%%%%%%%%%%%%%%%%%%%%%%

\section{PHYSICS RESULTS}
\label{sec:Physics}

The results of our maximum likelihood fits are summarized in 
Table~\ref{tab:results}. The statistical significance of the 
$B^0\to\rho^+\rho^-$ signal is 5.9~$\sigma$.
We find that the decay amplitude is predominantly longitudinal. 
To compute the branching fraction, 
we assume equal production rates for $\BzBzb$ and $\BpBm$.
To check the stability of our results we refit removing
each variable from the fit and find consistent results.
The number of fitted events, statistical significance,
branching fraction and polarization measurement errors, 
and the ML fit $\chi^2$ value are well reproduced with 
generated MC samples.

\begin{figure}[htbp]
\begin{center}
\vspace{1cm} 
\centerline{
\epsfig{figure=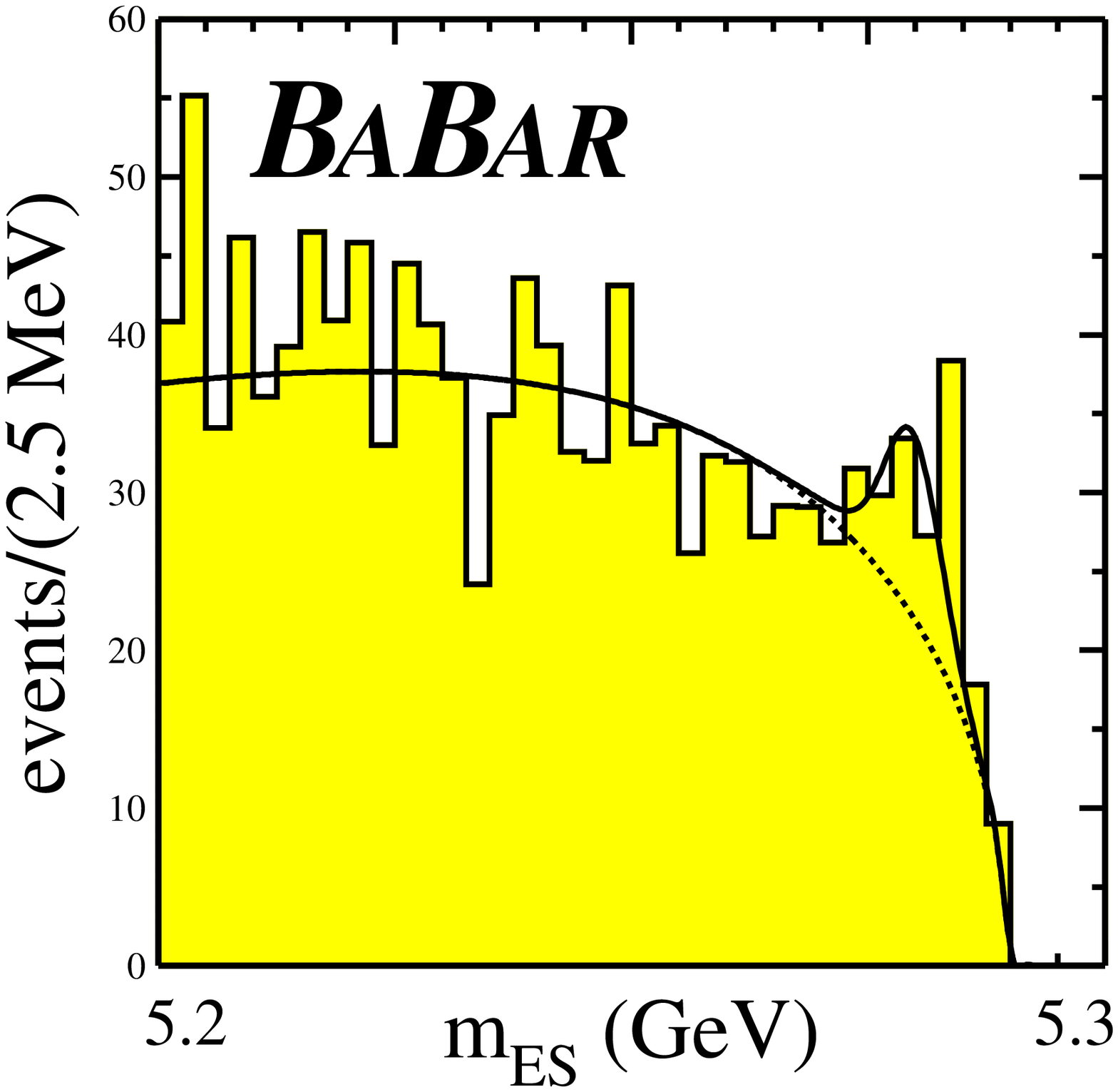,height=2.3in,width=3in}
\epsfig{figure=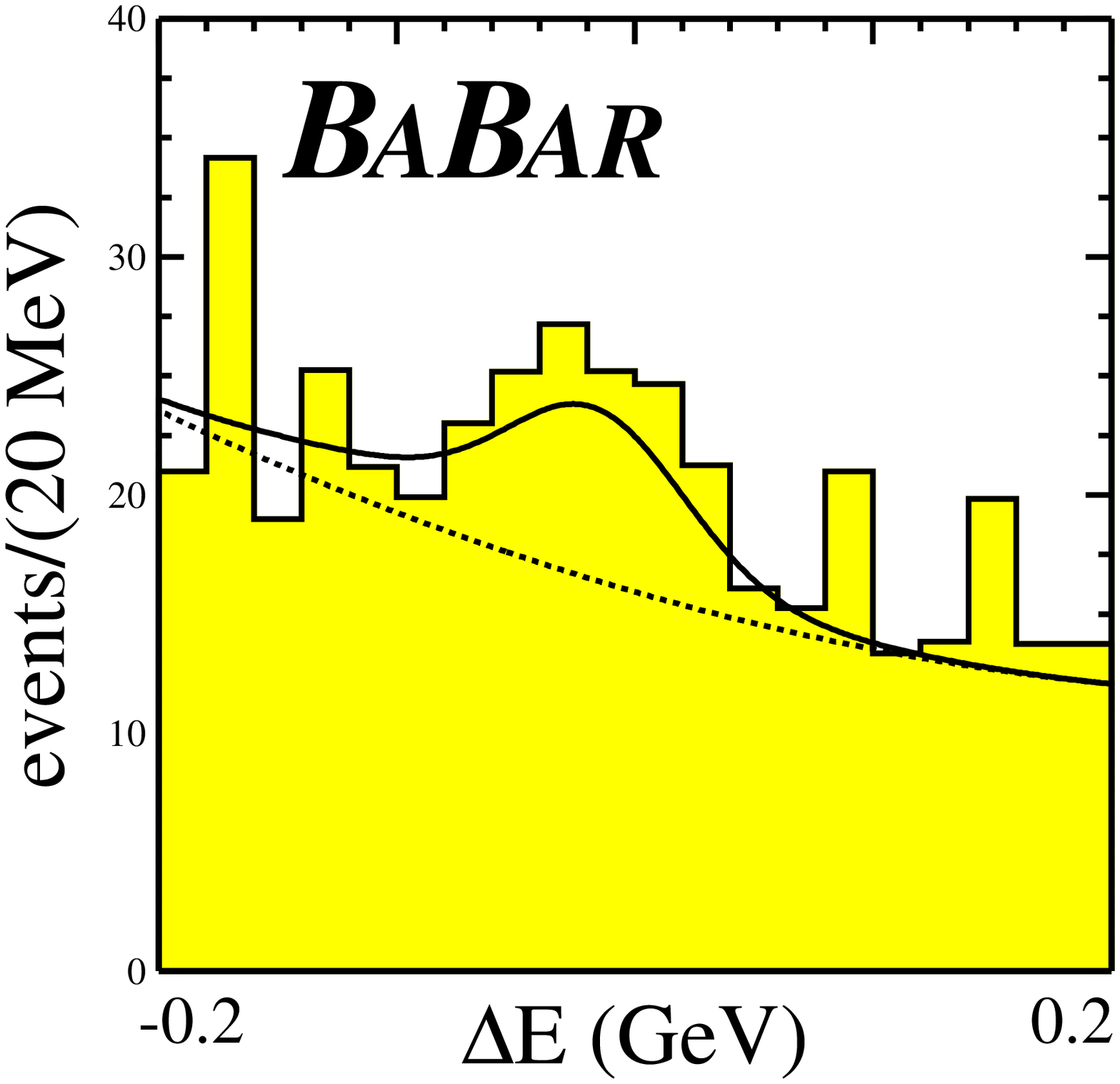,height=2.3in,width=3in}
}
\centerline{
\epsfig{figure=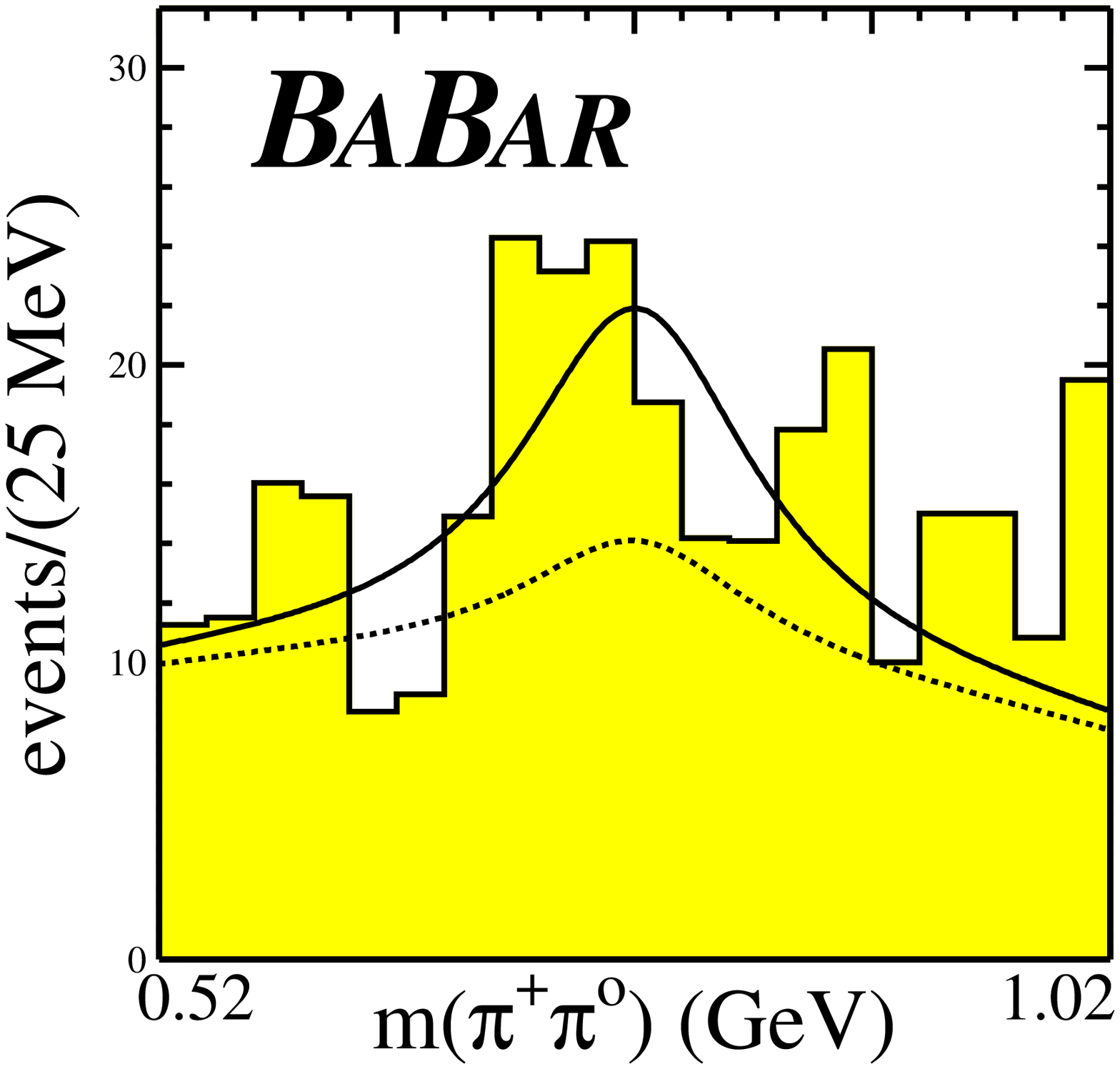,height=2.3in,width=3in}
\epsfig{figure=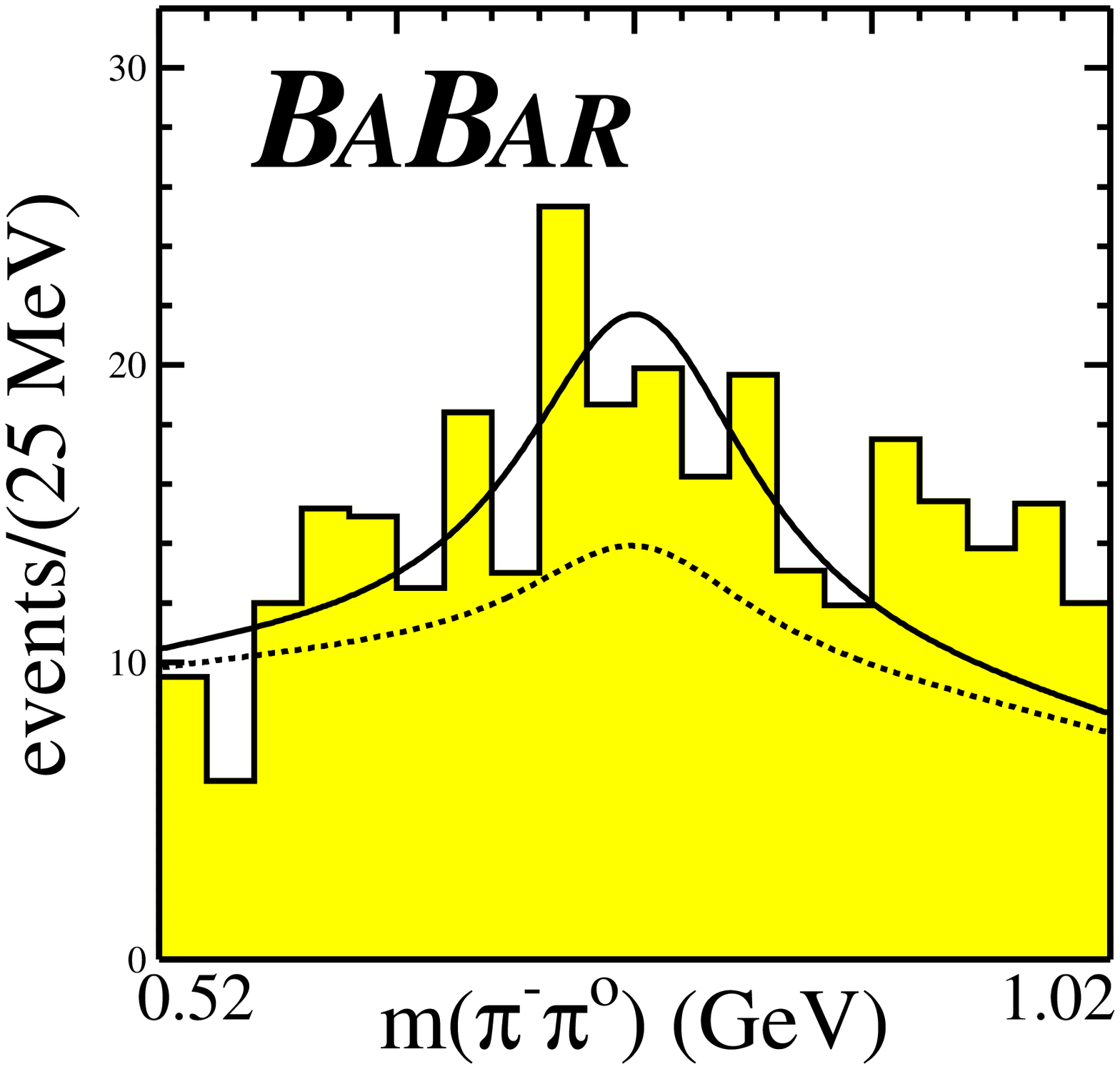,height=2.3in,width=3in}
}
\centerline{
\epsfig{figure=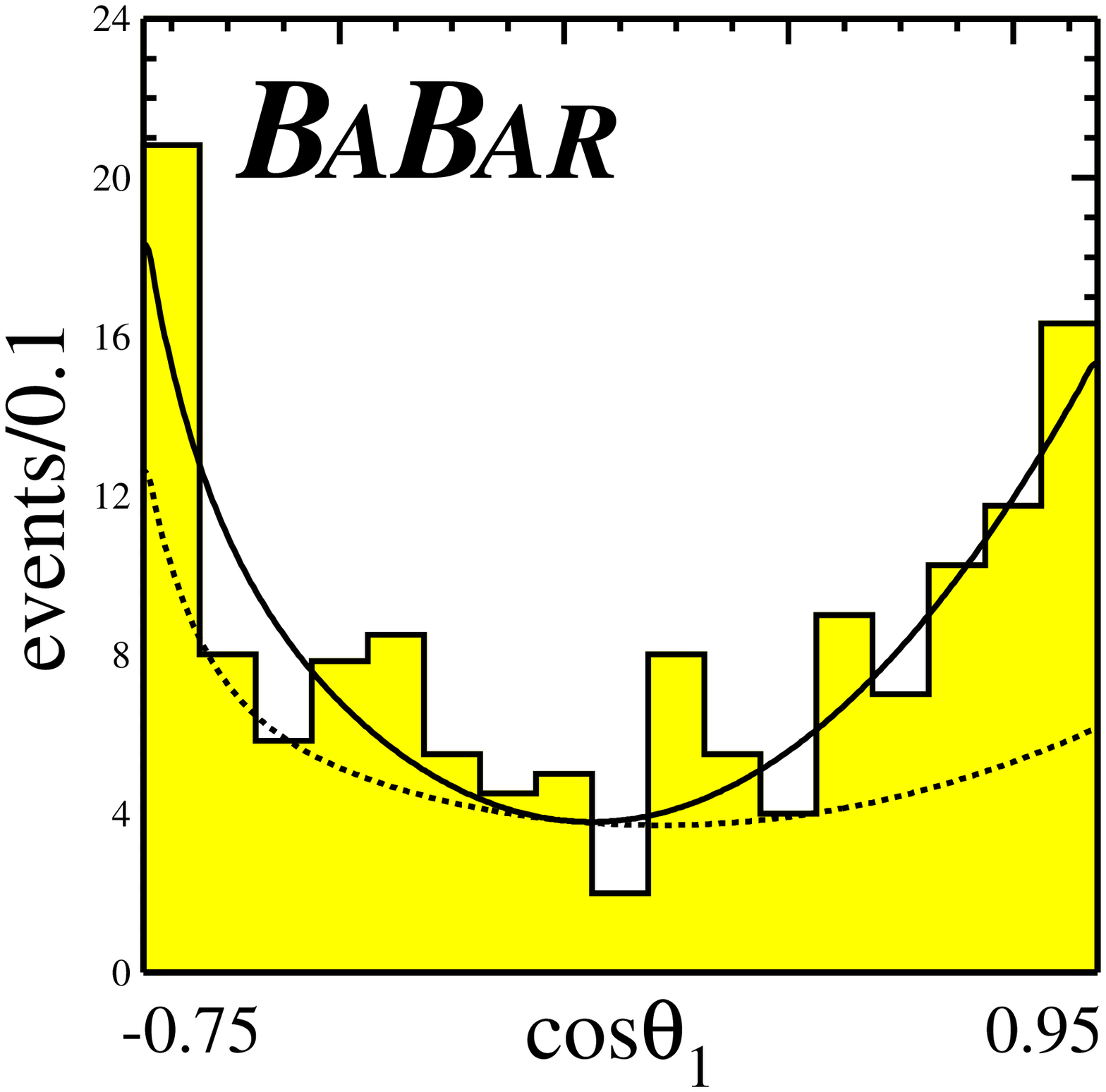,height=2.3in,width=3in}
\epsfig{figure=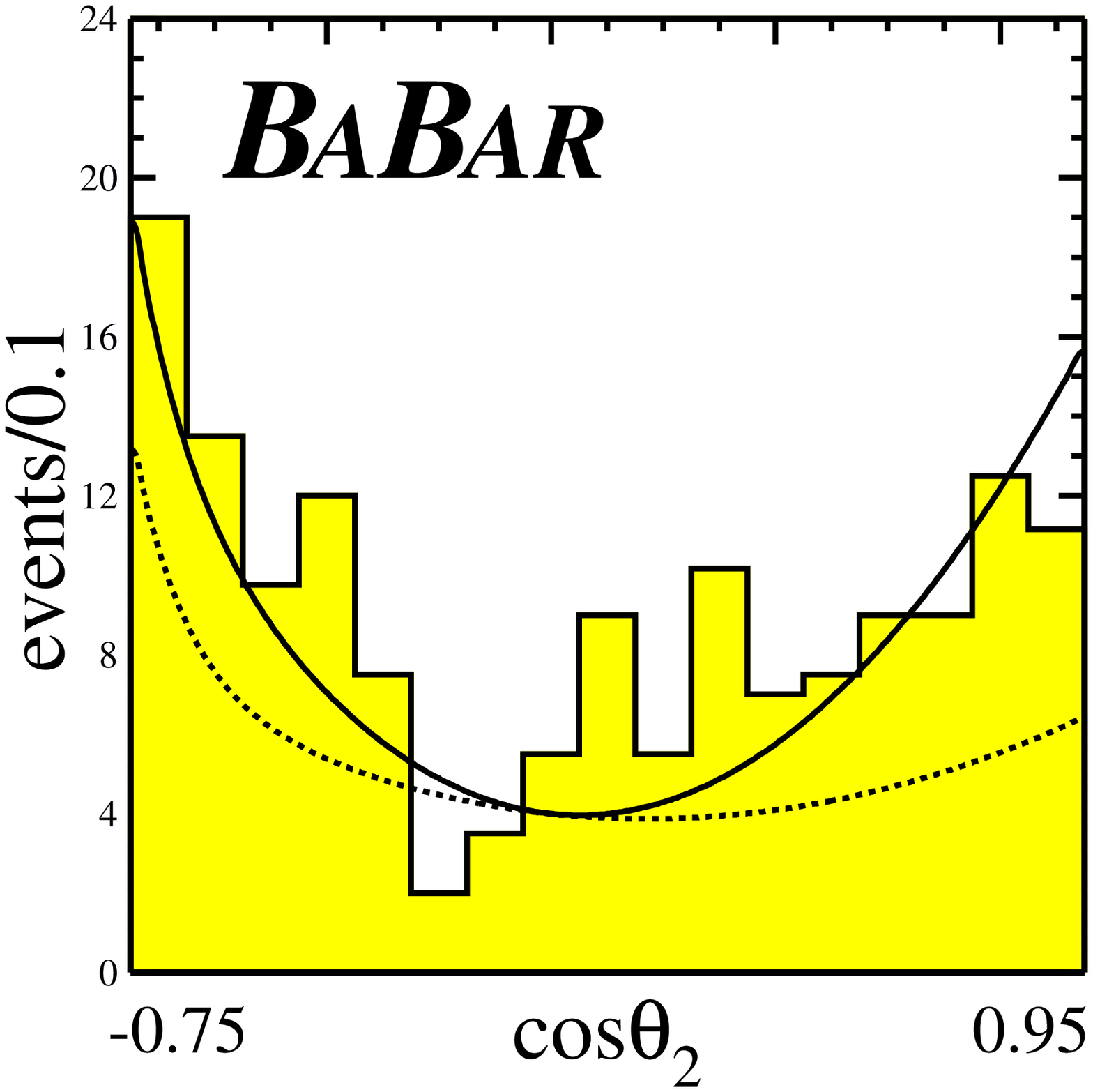,height=2.3in,width=3in}
}
\vspace{0.5cm}
\caption{\sl 
Projections onto the variables 
$m_{\rm{ES}}$, $\Delta E$,
$m_{\pi^+\pi^0}$, $m_{\pi^-\pi^0}$, 
$\cos\theta_1$, and $\cos\theta_2$ 
after a requirement on 
the signal-to-background probability ratio
${\cal P}_{\rm{sig}}/{\cal P}_{\rm{bkg}}$ with the 
PDF for the plotted variable excluded.
The histograms show the data, the solid (dashed) 
line shows the signal-plus-background 
(background only) PDF projection.
}
\label{fig:proj}
\end{center}
\end{figure}

The projections of the fit input variables
are shown in Fig.~\ref{fig:proj}.
The projections are made after a requirement on 
the signal-to-background probability ratio
${\cal P}_{\rm{sig}}/{\cal P}_{\rm{bkg}}$, where 
${\cal P}_{\rm{sig}}$ and ${\cal P}_{\rm{bkg}}$
are the signal and background PDFs defined in
Eq.~(\ref{eq:likelev}), but with the PDF for the 
plotted variable excluded. 
The histograms show the data with about 40$-$60$\%$
of signal retained, the lines show the PDF projections 
from the full sample.
Both mass and helicity-angle projections 
provide no evidence for non-resonant $B$ decays
with the same four-pion final states.

To check the sensitivity of our results to the 
presence of non-resonant $B^0\to\rho^\pm\pi^\mp\pi^0$
and $B^0\to\pi^+\pi^-\pi^0\pi^0$ decays, we
explicitly include a fit component assuming 
phase-space decay model. The selection requirements
alone suppress the $B\to\rho\pi\pi$ ($4\pi$) 
efficiency by one (two) order(s) of magnitude 
relative to $B^0\to\rho^+\rho^-$. 
The fit results with non-resonant component are consistent 
with our assumption of negligible non-resonant contribution. 
We exclude the hypothesis that all the signal is non-resonant 
$B\to\rho\pi\pi$ ($4\pi$) with $4.7$~$\sigma$ ($5.4$~$\sigma$)
statistical significance. 
However, we cannot exclude that $B\to\rho\pi\pi$ signal 
could represent $(9\pm15)\%$ (statistical errors only)
of our nominal $B^0\to\rho^+\rho^-$ event yield, where
we ignore interference effects. 

%%%%%%%%%%%%%%%%%%%%%%%%%%%%%%%%%%%%%%
\begin{table}[htbp]
\caption{\sl A summary of the fit results. The efficiency includes 
systematic errors, the significance with
systematic uncertainties is quoted in parentheses, 
while for other results, the first error
is statistical and the second systematic.}
\label{tab:results}
\begin{center}
\begin{tabular}{|c|c|}
\hline
\vspace{-2mm}
& \\
Reconstruction efficiency & $3.9_{-0.6}^{+0.9}~\%$ \\
\vspace{-2.5mm}&\\
Signal event yield ($n_{\rm sig}$)  & $93^{+23}_{-21}\pm 9$ \\
\vspace{-2.5mm}&\\
Statistical significance & 5.9 $\sigma$ (5.3 $\sigma$)  \\
\vspace{-2.5mm}&\\
Branching fraction (${\cal B}$) & $(27^{~+7~+5}_{~-6~-7})\times{10^{-6}}$ \\
\vspace{-2.5mm}&\\
\vspace{-2mm}
Signal polarization ($f_L$) & $0.99^{~+0.01}_{~-0.07}\pm 0.03$ \\
& \\
\hline
\end{tabular}
\end{center}
\end{table}
%%%%%%%%%%%%%%%%%%%%%%%%%%%%%%%%%%%%%%

%%%%%%%%%%%%%%%%%%%%%%%%%%%%%%%%%%%%%%
%%%%%%%%%%%%%%%%%%%%%%%%%%%%%%%%%%%%%%

%%%%%%%%%%%%%%%%%%%%%%%%%%%%%%%%%%%%%%%%%%%%%%%%%%%%%%%%%%%%%%%%%%%%%%%%%
% SYSTEMATIC UNCERTAINTIES
%%%%%%%%%%%%%%%%%%%%%%%%%%%%%%%%%%%%%%%%%%%%%%%%%%%%%%%%%%%%%%%%%%%%%%%%%

\section{SYSTEMATIC STUDIES}
\label{sec:Systematics}

Systematic uncertainties in the ML fit originate from assumptions 
about the PDF parameters.
Uncertainties in the PDF parameters arise from the limited
statistics in the background sideband data and signal control samples.
We vary them within their respective uncertainties,
and derive the associated systematic error on the event yield (9\%).
The signal remains statistically significant under
these variations ($5.3$~$\sigma$ including systematics).
Additional systematic errors in the number
of signal events originate from the uncertainty in the 
other $B$-decay cross-feed (3\%) that was studied with 
exclusive MC generated samples.

The systematic errors in the efficiency are due to track finding 
(2\% for two tracks), particle identification (2\% for two tracks), 
and $\pi^0$ reconstruction (13\% for two $\pi^0$s). 
The efficiency in the ML fit to signal samples, calculated
as the ratio of the fit signal yield over the
number of fully reconstructed decays in the fit sample, 
is less than 100\% because of fake combinations passing 
the selection criteria. 
We account for this in the efficiency evaluation and assign 
a systematic uncertainty of 7\% taken to be 1/2 of the 
inefficiency. 
The reconstruction efficiency depends on the decay 
polarization. We calculate the efficiencies using 
the measured polarization 
and assign a systematic error (${}_{~-2}^{+17}\%$)
corresponding to the total polarization measurement error.
Smaller systematic uncertainties arise
from event-selection criteria, MC statistics, 
and the number of produced $B$ mesons.

For the polarization measurement, 
we also include systematic errors from PDF variations 
that account for uncertainties in the detector acceptance and 
background parameterizations (0.025).
The biases from the resolution in helicity-angle measurement 
and dilution due to the presence of the fake combinations are
studied with MC simulation and are accounted for with a 
systematic error of 0.02.

%%%%%%%%%%%%%%%%%%%%%%%%%%%%%%%%%%%%%%%%%%%%%%%%%%%%%%%%%%%%%%%%%%%%%%%%%
% SUMMARY
%%%%%%%%%%%%%%%%%%%%%%%%%%%%%%%%%%%%%%%%%%%%%%%%%%%%%%%%%%%%%%%%%%%%%%%%%

\section{SUMMARY AND DISCUSSION}
\label{sec:Summary}

We have observed the decay $B^0\to\rho^+\rho^-$,
measured its branching fraction
${\cal B}=(27^{+7+5}_{-6-7})\times 10^{-6}$
and determined the longitudinal polarization fraction
${f_L}=0.99^{+0.01}_{-0.07}\pm 0.03$. 
Our results are preliminary. This completes 
the measurements of the isospin-related  
$B\to\rho\rho$ modes~\cite{babarVV, belle}.
The measured branching fraction agrees well with the more 
recent predictions in the range of 
$(18\hbox{--}35)\times 10^{-6}$~\cite{bvv3},
and the suppression of the transverse amplitude 
(by a factor of $m_\rho/m_B$) was expected~\cite{bvv3}.
These measurements improve our understanding of the
dynamics of hadronic weak decays and allow 
experimental tests of effective theories and
factorization~\cite{bvv1, bvv2, bvv3}.

The rates of the $B^0\to\rho^+\rho^-$ and $B^+\to\rho^0\rho^+$ 
decays appear to be larger than 
the corresponding rates of $B\to\pi\pi$ decays~\cite{pdg},
while the recent measurement of the $B^+\to\rho^0K^{*+}$ branching 
fraction~\cite{babarVV} does not show significant
enhancement with respect to $B\to\pi K$ decays~\cite{pdg}.
This indicates that the penguin pollution in the $B\to\rho\rho$ 
decay is smaller than in the $\pi\pi$ case,  
as predicted prior to the experimental measurements 
of the $B\to\rho\rho$ modes~\cite{bvv2}.
The dominance of the $\CP$-even longitudinal 
polarization in both $B^0\to\rho^+\rho^-$ and $B^+\to\rho^0\rho^+$
decays will also simplify $\CP$-violation studies.

At the same time, 
a relatively smaller experimental limit on the decay rate
can be achieved in the $B^0\to\rho^0\rho^0$ decays
than in the $B^0\to\pi^0\pi^0$ decays, this allows better 
experimental limit on the amount of penguin pollution 
in the $B\to\rho\rho$ decay amplitudes.
Since the tree contribution for the $B^0\to\rho^0\rho^0$ 
decay is color-suppressed, 
the decay rate is sensitive to the penguin diagram in
Fig.~\ref{fig:Diagram}.
Using the earlier $\babar$ measurements~\cite{babarVV} we obtain 
a 90\% confidence level upper limit on the ratio of the
longitudinal amplitudes $A_L$ in the $B\to\rho\rho$ decays:
%%%%%%%%%%%%%%%%%%%%%%%
\begin{equation}
\frac{|{A_L}(B^0\to\rho^0\rho^0)|^2+|{A_L}(\Bbar^0\to\rho^0\rho^0)|^2}
     {2\times|{A_L}(B^+\to\rho^0\rho^+)|^2}\equiv
\frac{{\cal B}(B^0\to\rho^0\rho^0)\times f_L(B^0\to\rho^0\rho^0)}
     {{\cal B}(B^+\to\rho^0\rho^+)\times f_L(B^+\to\rho^0\rho^+)}
< 0.10 \, .
\label{eq:rhorhoratio}
\end{equation}
%%%%%%%%%%%%%%%%%%%%%%%
In the above calculation
we assume conservatively the $B^0\to\rho^0\rho^0$ decay
polarization to be fully longitudinal ($f_L=1$),
use the average branching fraction measurements
for the $B$ and $\Bbar$ decays, and assume 
$|{A_L}(B^+\to\rho^0\rho^+)|=|{A_L}(B^-\to\rho^0\rho^-)|$
with only a tree-diagram contribution.
A similar experimental limit for the $B^0\to\pi^0\pi^0$ and 
$B^+\to\pi^+\pi^0$ decay amplitudes is 0.61~\cite{babarpi0pi0}, 
though still restrictive on the penguin pollution in 
the measurement of~$\alpha$~\cite{grossmanquinn}.
The above observations make the 
$B^0\to\rho^+\rho^-$ decay a promising channel to study $\CP$-violation
and set constraints on the weak-interaction angle $\alpha$.

%%%%%%%%%%%%%%%%%%%%%%%%%%%%%%%%%%%%%%%%%%%%%%%%%%%%%%%%%%%%%%%%%%%%%%%%%
% ACKNOWLEDGMENTS
%%%%%%%%%%%%%%%%%%%%%%%%%%%%%%%%%%%%%%%%%%%%%%%%%%%%%%%%%%%%%%%%%%%%%%%%%

\section{ACKNOWLEDGMENTS}
\label{sec:Acknowledgments}

We are grateful for the 
extraordinary contributions of our \pep2\ colleagues in
achieving the excellent luminosity and machine conditions
that have made this work possible.
The success of this project also relies critically on the 
expertise and dedication of the computing organizations that 
support \babar.
The collaborating institutions wish to thank 
SLAC for its support and the kind hospitality extended to them. 
This work is supported by the
US Department of Energy
and National Science Foundation, the
Natural Sciences and Engineering Research Council (Canada),
Institute of High Energy Physics (China), the
Commissariat \`a l'Energie Atomique and
Institut National de Physique Nucl\'eaire et de Physique des Particules
(France), the
Bundesministerium f\"ur Bildung und Forschung and
Deutsche Forschungsgemeinschaft
(Germany), the
Istituto Nazionale di Fisica Nucleare (Italy),
the Foundation for Fundamental Research on Matter (The Netherlands),
the Research Council of Norway, the
Ministry of Science and Technology of the Russian Federation, and the
Particle Physics and Astronomy Research Council (United Kingdom). 
Individuals have received support from 
the A. P. Sloan Foundation, 
the Research Corporation,
and the Alexander von Humboldt Foundation.

%%%%%%%%%%%%%%%%%%%%%%%%%%%%%%%%%%%%%%%%%%%%%%%%%%%%%%%%%%%%%%%%%%%%%%%%%
% BIBLIOGRAPHY
%%%%%%%%%%%%%%%%%%%%%%%%%%%%%%%%%%%%%%%%%%%%%%%%%%%%%%%%%%%%%%%%%%%%%%%%%


\begin{thebibliography}{99}

\bibitem{Kobayashi}
\label{ref:Kobayashi}
M.~Kobayashi and T. Maskawa, Prog. Theor. Phys. {\bf 49}, 652 (1973).

\bibitem{bvv1}
G.~Kramer and W.F.~Palmer, 
Phys.\ Rev.\ D {\bf 45}, 193 (1992).

\bibitem{bvv2}
R.~Aleksan {\it et al.}, 
Phys.\ Lett.\ B {\bf 356}, 95 (1995).

\bibitem{bvv3}
D.~Ebert, R.N. Faustov, and V.O.~Galkin,
Phys. Rev. D {\bf 56}, 312 (1997);
A.~Ali, G.~Kramer, and C.-D.~Lu,
Phys. Rev. D {\bf 58}, 094009 (1998);
Y.-H.~Chen {\it et al.},
Phys. Rev. D {\bf 60}, 094014 (1999);
H.-Y.~Cheng and K.-C.~Yang,
Phys.\ Lett.\ B {\bf 511}, 40 (2001).

\bibitem{grossmanquinn}
M.~Gronau and D.~London, Phys.\ Rev.\ Lett. {\bf 65}, 3381 (1990);
Y.~Grossman and H.~Quinn, Phys.\ Rev.\ D {\bf 58}, 017504 (1998).

\bibitem{babarVV}
\babar\ Collaboration, B.~Aubert {\it et al.},
BABAR-PUB-03-018, hep-ex/0307026,
submitted to Phys.\ Rev.\ Lett.

\bibitem{belle}
BELLE Collaboration, J. Zhang {\it et al.},
BELLE-2003-6, hep-ex/0306007,
submitted to Phys.\ Rev.\ Lett.

\bibitem{babar}
\babar\ Collaboration, B.~Aubert {\it et al.},
Nucl.\ Instrum.\ Methods {\bf A479}, 1 (2002).

\bibitem{pep} 
PEP-II Conceptual Design Report, SLAC-R-418 (1993).

\bibitem{CLEO-fisher}
CLEO Collaboration,
D.M.~Asner {\it et al.}, 
Phys.\ Rev.\ D {\bf 53}, 1039 (1996).

\bibitem{geant}
The \babar\ detector Monte Carlo 
simulation is based on GEANT:
R.~Brun {\it et al.}, CERN DD/EE/84-1.

\bibitem{argus}
ARGUS Collaboration, H.~Albrecht {\it et al.}, Phys.\ Lett.\ B {\bf 241}, 278 (1990).

\bibitem{pdg} 
Particle Data Group,
K. Hagiwara {\it et al.}, Phys.\ Rev.\ D {\bf 66}, 010001 (2002).

\bibitem{babarpi0pi0}
\babar\ Collaboration, B.~Aubert {\it et al.},
Phys.\ Rev.\ Lett. {\bf 91}, 021801 (2003).

\end{thebibliography}
\end{document}